\documentclass[solaromanenum]{solarphysics}
\usepackage[hyperref,optionalrh,solaromanenum]{spr-sola-addons} 
\usepackage{graphicx}                    
\usepackage{color}                       
\usepackage{breakurl}                         
\usepackage{overpic}

\begin{document}
\begin{article}
\begin{opening}
\title{Detections of Multi-Periodic Oscillations during a Circular Ribbon Flare}


\author[addressref={aff1,aff2},email={ningzongjun@pmo.ac.cn}]{\inits{Z. J}\fnm{Zongjun}~\lnm{Ning}\orcid{0000-0002-9893-4711}}
\author[addressref={aff1},email={}]{\inits{ }\fnm{Ya}~\lnm{Wang}\orcid{0000-0003-3699-4986}}
\author[addressref={aff1,aff2},email={}]{\inits{ }\fnm{Zhenxiang}~\lnm{Hong}}
\author[addressref={aff1},corref,email={lidong@pmo.ac.cn}]{\inits{D.}\fnm{Dong}~\lnm{Li}\orcid{0000-0002-4538-9350}}
\runningauthor{Z. Ning et al.}
\runningtitle{Detections of Multi-periodic Oscillations During a Circular Ribbon Flare}

\address[id=aff1]{Key Laboratory of Dark Matter and Space Astronomy, Purple Mountain Observatory, CAS, Nanjing 210023, China}
\address[id=aff2]{School of Astronomy and Space Science, University of Science and Technology of China, Hefei 230026, China}
\begin{abstract}
We present the analysis of three kinds of oscillating behavior using
multi-wavelength observations of the 10 November 2013
(SOL2013-11-10T05:14) circular-ribbon flare. This event is a typical
circular-ribbon flare with an outer spine structure and homologous
jets. We found three kinds of oscillations (or perturbations): i)
flux oscillation (or QPP) with a dominant period of about 20\,seconds
in X-ray, EUV, and microwave emissions, ii) periodic jets with an
intermittent cadence of around 72\,seconds, iii) an outer loop perturbing
half a cycle with a duration of about 168\,seconds. Similar to the
periodic jets that could be produced by a nonthermal process, like
repeated magnetic reconnection, the flare QPP detected in the thermal
emissions could have the same origin as the oscillation seen in
the nonthermal emissions. The outer-loop perturbation is possibly
triggered by a blast wave driven by the circular-ribbon flare, or it
might be modulated by the sausage wave or the slow magnetoacoustic
wave. The results obtained provide data for further numerical studies on
the physical origin of the flare oscillations.
\end{abstract}


\end{opening}
\section{Introduction}
A solar flare is a powerful and impulsive process of
magnetic-free-energy release in the solar atmosphere \citep[see][for
a recent review]{Benz17}. Solar-flare topology often shows elongated
structures seen in the H$\alpha$, white light, and ultraviolet (UV)
wavelengths, called ``flare ribbons''. Typically, two ribbons are
located in regions of opposite magnetic polarities, which are
parallel to the magnetic polarity inversion line (PIL) and separating
the polarities. This is known as a `` two-ribbon flare'' (a
detailed sketch can be seen in \cite{Priest02}) and could be well
explained by the classical 2D-reconnection model
\citep{Sturrock64,Shibata11,Yan18}. However, a solar flare actually
occurs in a more complicated structure associated with the 3D
magnetic null point, i.e. the fan--spine topology \citep{Torok09}.
The null point is mainly passed through by the inner and outer spine
field lines, and the closed separatrix surface is dominated by the
fan structure with a dome-like shape. The magnetic fields at the
inner spine footpoint are opposite to those of the dome-shaped fan,
producing a circular PIL. This eruptive feature is called a
``circular-ribbon flare'' \citep{Masson09,Wang12}, which is regarded
as a result of the null-point reconnection. A schematic picture
illustrating the main components and geometrical structures of the
circular-ribbon flare was provided by \cite{Wang12}. During its
eruption, the flare radiation at the footpoint of the fan field lines is
largely constituting a closed circular ribbon, and at the footpoints
of the inner spine field lines is a compact source called a
central/inner ribbon. On the other hand, the outer spine field lines
could be closed or open, and a remote/outer ribbon is formed at the
footpoint of the closed outer spine field lines, while a series of
homologous jets could be successively produced along the
magnetic-field lines of the open outer spine
\citep[e.g.][]{Chifor08,Liu11,Reid12,Hao17,Hernandez17,Lit18,Song18,Zhang21}.
This combination of factors results in various kinds of oscillations
relating to this type of flare
\citep{Meszarosova13,Kumar15,Kashapova20,Zhang20}.

Quasi-periodic pulsations (QPPs) are intensity oscillations during
solar flares \citep[e.g.][]{Van16,McLaughlin18,Zimovets21}. They
often appear as a series of regular and periodic pulses in the flare's
light curve, which could be detected in a broad range of
wavelengths, i.e. radio/microwave emissions
\citep{Ning05,Kupriyanova10,Nakariakov18,Karlicky21}, H\,$\alpha$
\citep{Srivastava08,Yang16,Kashapova20,Li20a} and Ly\,$\alpha$
\citep{Milligan17,Li20b,Li21,Lu21} emissions, UV or extreme-UV (EUV)
wavelengths \citep{Pugh17,Shen18,Kobanov19,Yuan19}, soft/hard X-ray
(SXR/HXR) channels \citep{Ning14,Inglis16,Li17,Kolotkov18,Hayes20},
and even $\gamma$-ray emission \citep{Nakariakov10,Li20c}. Such
oscillatory features were also observed in the circular-ribbon
flare \citep{Meszarosova13,Kumar15,Kashapova20}. A typical flare QPP
is generally characterized by repetition and periodicity, and the
quasi-periods can be observed from sub-seconds through seconds and
minutes to dozens of minutes
\citep{Fleishman02,Tan10,Dolla12,Kolotkov15,Kupriyanova16,Ning17,Yu19,Clarke21,Li21a}.
It seems that the various periods of flare QPPs are always dependent
on the detected wavelengths or the temporal resolution of the
measuring instruments \citep[see][for a recent
review]{Kupriyanova20}. These properties indicate the possibility of
generating flare QPPs with various periods by different mechanisms
\citep[e.g.][]{McLaughlin18,Kupriyanova20}. The short-periodicity
QPPs (i.e. sub-seconds) are generally observed in radio or HXR
emissions due to their high temporal resolution, which are most
likely associated with the kinetic process driven by the dynamical
interaction between electromagnetic plasmas and energetic particles
trapped in the flaring loop
\citep[e.g.][]{Aschwanden87,Fleishman02,Ning05,Tan10,Yu19}. The
flare QPPs with long periods on the order of seconds and minutes
can be discovered in a broad wavelength range, such as microwave,
white light, UV/EUV, SXR/HXR, and even $\gamma$-rays
\citep{Nakariakov10,Dolla12,Kolotkov15,Li17,Kobanov19,Li20a,Hayes20,Li21}.
They are probably related to magnetohydrodynamic (MHD) waves
in complicated magnetic configurations in the flare region or might
be associated with the solar global oscillations
\citep{Ofman02,Tian16,Kupriyanova16,Shen18,Kolotkov18,Yuan19,Nakariakov20,Clarke21}.

Flare QPPs have attracted a lot of attention since they were first
found in solar X-ray emissions \citep{Parks69}. However, the
physical mechanism responsible for the generation of flare QPPs is
still an open question \citep{Van16,McLaughlin18,Kupriyanova20}. They
are generally attributed to MHD waves in slow mode, kink mode, or
sausage mode
\citep{Yuan16,Lib20,Nakariakov20,Nakariakov21,Amiri21,Wang21}. They
might also be caused by repetitive magnetic reconnections, which
could accelerate nonthermal electrons and produce flare emissions
with quasi-periodic pulses
\citep[e.g.][]{Kliem00,Chen06,Nakariakov11,Kumar15,Lit15,Guidoni16,Yuan19,Li20c}.
On the other hand, flare QPPs are also interpreted as a
self-oscillating processes, such as magnetic dripping
\citep{Nakariakov10b,Li20b} or may be explained by the magnetic
tuning-fork model \citep{Takasao16} and the LRC-circuit model
\citep{Tan16,Chen19,Li20a}. The supra-arcade downflows were found to
collide with post-flare loops at different instances, which could
lead to the quasi-periodicity of the flare radiation at wavelengths of
UV/EUV and X-ray \citep{Cai19,Xue20,Samanta21}. Moreover, the flare
QPPs are thought to be related to the fundamental physical processes
of solar flares, i.e. magnetic reconnection, energy release,
particle acceleration, and also MHD waves
\citep[e.g.][]{Nakariakov11}. Therefore, they are useful for
remotely diagnosing the key physical parameters in flare regions
\citep{Brosius15,Dominique18,Yuan19}.

Thanks to high-resolution imaging telescopes in EUV/SXR channels,
the spatial motions of displacements have been found in hot coronal
loops, and they are regarded as coronal-loop oscillations, which are
often related to solar flares
\citep[e.g.][]{Nakariakov99,Aschwanden02}. Coronal-loop
oscillations are also seen as Doppler-shift oscillations using
spectroscopic observations \citep{Wang02,Tian16,Li18}. Their
displacement amplitudes are found to range from sub-megameters to
dozens of megameters
\citep{Anfinogentov15,Su18,Goddard20,Nakariakov21}, and their
oscillation periods are strongly dependent on loop lengths.
While the ratio of the decay time to the oscillation period is
systematically related with the displacement amplitudes
\citep{Goddard16}. Fast magnetoacoustic waves, detected as wavelet
tadpoles, were found in the fan structure of a circular-ribbon
flare \citep{Meszarosova13}. Recently, transverse oscillations were discovered in the outer loop of a circular-ribbon flare, which
are explained as kink modes of the MHD wave
\citep{Zhang20b,Dai21}, largely due to their incompressibility or
weak compressibility \citep{Yuan16,Amiri21}. These circular-ribbon
flares were also found to be accompanied by repeated and homologous
coronal jets. On the other hand, the transverse oscillations
detected in the coronal loops can also be interpreted as sausage-mode
waves in the plasma loops, and they are often caused by density
perturbations in the coronal loops \citep{Tian16,Lim18,Lib20}.
However, sausage oscillations excited by circular-ribbon flares
are rarely reported in the outer coronal loop.

In this study, we investigated three kinds of oscillating behavior
using multi-wavelength observations; they were all related to a
circular-ribbon flare on 10 November 2013. The article is organized
as follows: Section 2 introduces the observations and instruments,
Section 3 presents the data analysis and our main results, and
Section 4 gives the conclusion and discussion.

\section{Observations and Instruments}
We have studied the solar flare of 10 November 2013 (SOL2013-11-10T05:14). It was an X1.1 class
flare and occurred in the active region NOAA~11890 (S11W28). The
event started at about 05:09~UT, reached its maximum at around 05:14~UT,
and ended at roughly 05:22~UT. The flare was simultaneously observed
in multiple wavelengths, which was measured by the {\it
Geostationary Operational Environmental Satellite} (GOES), the {\it
Reuven Ramaty High Energy Solar Spectroscopic Imager}
\citep[RHESSI:][]{Lin02}, the {\it Nobeyama RadioPolarimeters}
\citep[NoRP:][]{Nakajima85}, the {\it Nobeyama RadioHeliograph}
\citep[NoRH:][]{Hanaoka94}, {\it Konus-Wind}
\citep{Aptekar95,Palshin14}, the {\it Gamma Ray Burst Monitor} (GBM)
on {\it Fermi} \citep{Meegan09}, the {\it Atmospheric Imaging
Assembly} \citep[AIA:][]{Lemen12}, the {\it Helioseismic and
Magnetic Imager} \citep[HMI:][]{Schou12}, and the {\it Extreme
Ultraviolet Variability Experiment} \cite[EVE:][]{Woods12} onboard
the {\it Solar Dynamics Observatory} \citep[SDO:][]{Pesnell12}; see
the details in Table~\ref{tab1}.

\begin{table}
\caption{Instruments used to detect flare
oscillations in this study.} \setlength{\tabcolsep}{5pt}
\begin{tabular}{c c c c c c}
\hline
Instrument   &  Cadence      &    Channels         &  Description  & Observation    \\
\hline
             &               &    1\,--\,8\,\AA      &   SXR         &                \\
GOES/XRS     &  $\approx$2\,seconds    &    0.5\,--\,4\,\AA    &   SXR  &     1D         \\
\hline
             &               &     6\,--\,12\,keV       &   SXR         &                \\
RHESSI       &     4\,seconds    &    50\,--\,100\,keV  &   HXR         &    2D          \\
\hline
             &               &   4.2\,--\,11.5\,keV    &   SXR         &                 \\
{\it Fermi}/GBM    &  $\approx$0.256\,second &   11.5\,--\,26.6\,keV   &   SXR/HXR     &    1D          \\
             &               &   26.6\,--\,50.4\,keV   &   HXR         &                \\
             &               &   50.4\,--\,102.4\,keV  &   HXR         &                  \\
\hline
             &               &    20$-$80\,keV      &   SXR/HXR   &       \\
{\it Konus-Wind}   &  $\approx$1.024\,seconds &    80$-$300\,keV     &   HXR       & 1D     \\
             &               &    300$-$1200\,keV   &   HXR/$\gamma$-ray & \\
\hline
             &               &      1\,GHz          &   Microwave  &      \\
             &               &      2\,GHz          &   Microwave  &      \\
NoRP         &     0.1\,second     &     3.75\,GHz       &   Microwave  &  1D  \\
             &               &      9\,GHz          &   Microwave  &      \\
             &               &      17\,GHz         &   Microwave  &      \\
             &               &      35\,GHz         &   Microwave  &      \\
\hline
             &               &      17\,GHz         &   Microwave  &      \\
NoRH         &     1\,second       &      34\,GHz         &   Microwave  & 2D   \\
\hline
             &               &    0.1\,--\,7.0\,nm     &     SXR      &      \\
SDO/EVE      &    0.25\,second     &    17.2\,--\,20.6\,nm   &     EUV      & 1D   \\
             &               &    23.1\,--\,27.6\,nm   &     EUV      &      \\
             &               &    28.0\,--\,32.6\,nm   &     EUV      &      \\
\hline
             &    12\,seconds       &    171\,\AA        &     EUV      &      \\
SDO/AIA      &    24\,seconds       &    1600\,\AA       &     UV       &  2D  \\
\hline
SDO/HMI      &    45~seconds       &     magnetogram     &     LOS      &  2D  \\
\hline
\end{tabular}
\label{tab1}
\end{table}

\begin{figure}
\centerline{\includegraphics[width=\textwidth,clip=]{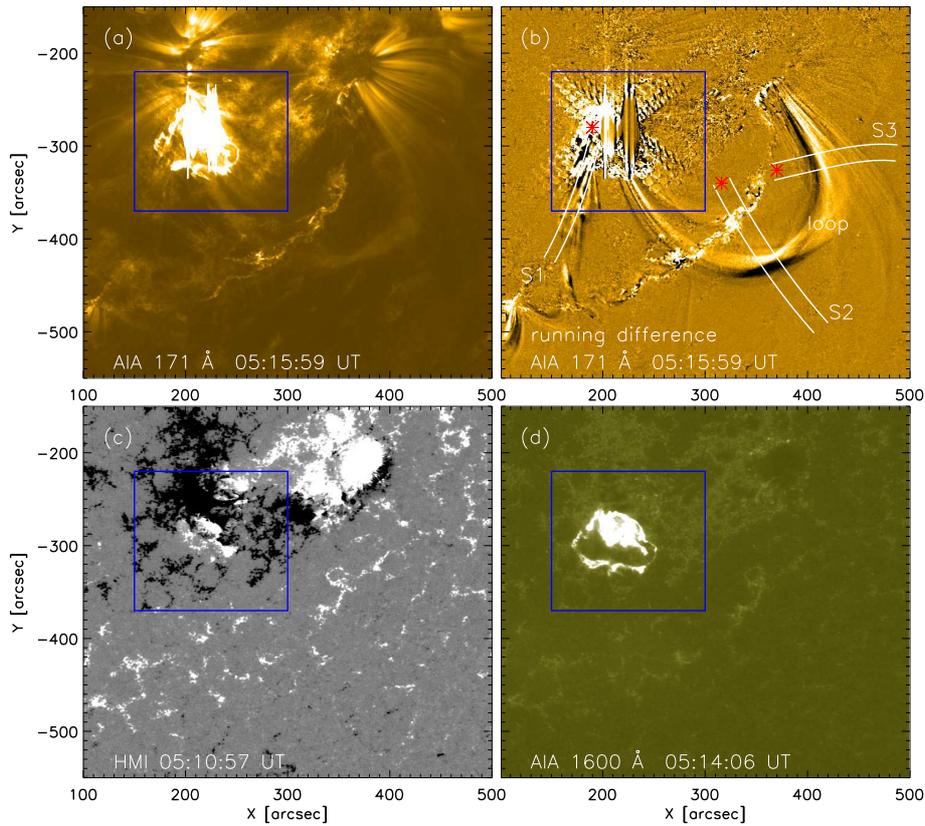}}
\caption{Panels~{\bf a} and {\bf b}: AIA~171\,\AA\ and its
running-difference images show the circular-ribbon flare with a FOV
of 400$^{\prime\prime}$$\times$400$^{\prime\prime}$. Three groups of
{\it solid lines} outline the slit positions in Figure~\ref{tds}, and the
{\it red symbols} of ``$*$'' indicate the zero of the $y$-axis.
Panels {\bf c} and {\bf d}: HMI LOS magnetogram and AIA~1600\,\AA\
image with a FOV of
400$^{\prime\prime}$$\times$400$^{\prime\prime}$. The {\it blue box}
outlines a small FOV of about
150$^{\prime\prime}$$\times$150$^{\prime\prime}$ shown in
Figure~\ref{extr}{\bf a} and {\bf b}.} \label{image}
\end{figure}

The X1.1 flare shows a closed circular ribbon and a faint remote
ribbon, as well as a faint outer loop that connected the closed
circular ribbon and a remote brightening, which is seen in
171\,\AA~and 1600\,\AA~images measured by SDO/AIA shown in
Figure~\ref{image}a, b, and d, and in the Electronic Supplementary
Material movie. The AIA 1600\,\AA~image was obtained at 05:14:16 UT,
which is around the flare maximum, while the AIA~171\,\AA~band image
was taken at 05:15:59~UT. At this moment, the outer loop can be more
clearly distinguished, especially in the running difference. The
lines marked S1\,--\,S3 show the positions of the slits on the outer
loop. They are selected to study the jets and the loop perturbation, which will be
analyzed in the next section, and the red ``$*$'' symbols mark the
zero of the $y$-axis in Figure~\ref{tds}. Figure~\ref{image}c gives
the line-of-sight(LOS) magnetogram obtained by SDO/HMI
\citep{Schou12}. A closed circular ribbon is associated with the
negative magnetic polarity and an inner ribbon is on the compact
positive magnetic polarity, while the faint remote ribbon is along
the diffuse positive magnetic polarity. The flare event was also
observed by the RHESSI and NoRH images in X-rays and microwaves, as
indicated by the color contours in Figure~\ref{extr}a and b. The SXR
and HXR images at 6\,--\,12\,keV and 50\,--\,100\,keV were
reconstructed using the CLEAN algorithm from the RHESSI data for the
time interval of one~minute. This processing allows improving the
signal-to-noise ratio. Similar to the AIA~1600\,\AA~band~image, the
microwave images are obtained at 05:14:16~UT. This event at SXR
6\,--\,12\,keV (blue contours) and microwave 17\,GHz (green contours)
and 34\,GHz (cyan contours) shows one source, but double HXR
footpoint sources at 50\,--\,100\,keV (red contours).
Both the SXR/HXR and microwave emissions appear to overlay the inner
flare ribbon seen in the AIA~1600\,\AA~band image. This is
consistent with the standard flare model
\citep[e.g.][]{Sturrock64,Shibata11}, for instance, the double
footpoint sources of HXR at 50\,--\,100\,keV (red contours) are
located in the regions with different polarity magnetic fields: one
is on the inner ribbon with negative polarity field, while the other
is on the circular ribbon with positive polarity.

The X-ray observations have a different temporal resolution than the
other instruments (see Table~1). It should be pointed out that the
X-ray light curves recorded by {\it Konus-Wind} and {\it Fermi}/GBM
have been interpolated to the uniform temporal resolution, i.e.
1.024\,seconds for the {\it Konus-Wind} data \citep[e.g.][]{Li20c}
and 0.256\,seconds for the {\it Fermi}/GBM data \citep[see
also,][]{Li15,Ning17}. NoRP is a ground-based telescope, which
provides solar microwave fluxes at six frequencies with a temporal
cadence of $\approx$0.1\,second \citep{Nakajima85}. The {\it
EUV SpectroPhotometer} for SDO/EVE measures the Sun in one SXR and
three EUV wavebands with a temporal resolution of 0.25\,seconds
\citep{Didkovsky12,Woods12}. Similar to the GOES~1\,--\,8\,{\AA}
light curve, the ESP~0.1\,--\,7\,nm is a SXR flux. The
ESP~17.2\,--\,20.6\,nm flux includes the line emission mostly from
the coronal plasma at about 1\,--\,2\,MK, while the
ESP~23.1\,--\,27.6\,nm and 28.0\,--\,32.6\,nm fluxes are mainly
dominated by the He {\sc ii} line in the chromosphere
\citep{Didkovsky12,Dolla12,Li21r}.

\section{Data Analysis and Results}
\subsection{NLFFF Extrapolation}
\begin{figure}
\centerline{\includegraphics[width=\textwidth,clip=]{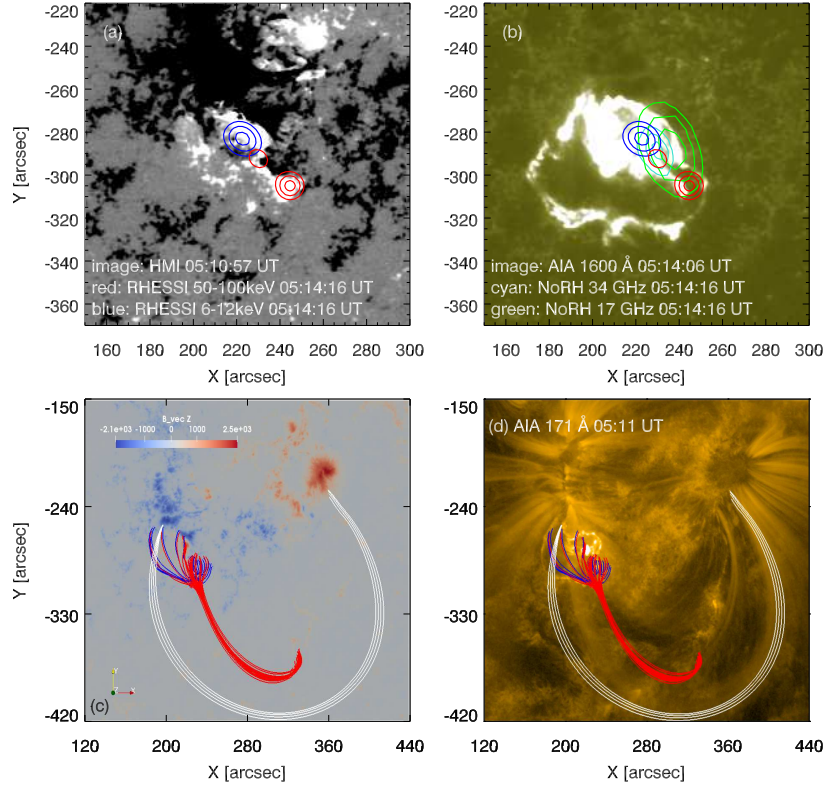}}
\caption{Panels~{\bf a} and {\bf b}: HMI LOS magnetogram and
AIA~1600\,\AA\ image with a small FOV. The {\it red and blue
contours} represent the RHESSI X-ray emissions at 50\,--\,100\,keV
and 6\,--\,12\,keV, while the {\it cyan and green contours} show the
microwave emissions in NORH~34\,GHz and 17\,GHz. The levels are set
at 50\,\%, 70\,\%, and 90\,\% of the maximum brightness of each
detected channel. Panel~{\bf c}: Magnetic topology for the flare
region in the $x-y$ plane observed from the $z$-direction. The $B_z$
component at the bottom boundary ranges from -2.1$\times$10$^3$ to
2.5$\times$10$^3$\,Gauss. Panel~{\bf d}: AIA 171\,\AA\ image at
$\approx$05:11 UT overlaid with the extrapolated magnetic-field
lines. The {\it red and blue lines} represent the spine--fan
magnetic topology, while the {\it white lines} indicate the
overlying magnetic loop.}\label{extr}
\end{figure}

To reveal the magnetic topology of the circular-ribbon flare region,
we carried out a nonlinear force-free field (NLFFF) extrapolation
using the optimization approach proposed by \cite{Wheatland00} and
further developed by \cite{Wiegelmann04}. We used the HMI vector
magnetogram retrieved from the HMI.B\_720\,seconds data series at
about 05:00:00~UT as the bottom boundary. The 180-degree ambiguity
in the transverse components of the vector magnetic field was removed
using the improved minimum energy method
\citep{Metcalf06,Leka09}, and the projection effect was also
corrected. Then, the vector magnetic field can be used as a
boundary condition after a preprocessing to satisfy the force-free
and torque-free conditions on the photosphere \citep{Guo17}. The
extrapolation cube size is 232$\times$196$\times$196\,Mm$^3$. As
shown in Figure~\ref{extr}c and d, the flare shows a spine--fan
magnetic topology, which is similar to the previous result about the
circular-ribbon flare \citep[e.g.][]{Meszarosova13}. During the
flare event, the dome-like structure is consistent with the closed
circular ribbon that is accompanied by the occurrence of a series of
coronal jets. This is also followed by the remote ribbon at the
western end of the outer spine. In addition, a set of overlying
magnetic-field lines indicated by white lines matches well with the
faint outer loop observed in AIA~171\,\AA. Consistent with the
previous observation \citep[e.g.][]{Wang12}, this circular-ribbon
flare has two HXR footpoint sources on the optical inner and circular
ribbons, respectively. The microwave sources at 17\,GHz (green
contours) and 34\,GHz (cyan contours) are located between two HXR footpoint
sources. It is reasonable that the microwave emissions at 17/34\,GHz
are produced by the electrons trapped in the flare loops (part of
the fan structure in Figure~\ref{extr}c and d) through synchrotron
radiation. The microwave source at 34\,GHz is smaller than that at
17\,GHz, because it comes from the higher electrons located near the
footpoint of the magnetic loop. The thermal source of SXR at
6\,--\,12\,keV (blue contours) is not located between two HXR footpoint
sources, but at the top of the inner ribbon. Except for the
projection effect, another possible location seems to be near the null
point of the magnetic field, as shown in Figure~\ref{extr} c and d.
The null point is thought to be the reconnection site of the
circular-ribbon flare, and is near the fan-structure top (the
loop apex in the standard flare model) where the SXRs radiate.

\subsection{Flux Oscillation During the Circular Ribbon Flare}
\begin{figure}
\centerline{\includegraphics[width=0.9\textwidth,clip=]{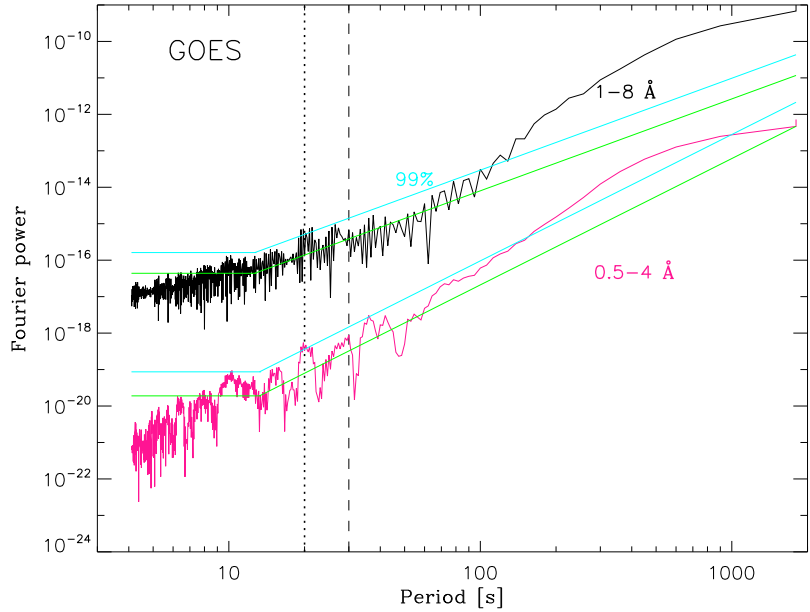}}
\caption{Fourier power of the flare light curves measured by the
GOES at 1\,--\,8\,\AA\ ({\it black}) and 0.5\,--\,4\,\AA\ ({\it
magenta}). The {\it vertical dotted and dashed lines} mark the positions
with a periodicity of 20\,seconds and 30\,seconds. The {\it green
lines} are the best fit, and the {\it cyan lines} represent the
99\,\% significance level.} \label{ffts}
\end{figure}

To investigate the periodicity of this circular-ribbon flare, we
first perform the Fast Fourier Transform (FFT) of the raw light
curves in SXR 1\,--\,8\,\AA\ (black) and 0.5\,--\,4\,\AA\ (magenta)
measured by the GOES, as shown in Figure~\ref{ffts}. Similar to
previous results
\citep[e.g.][]{Inglis16,Kolotkov18,Hayes20,Li20c,Wang20}, each
Fourier power spectrum of the raw light curve shows two dominant
features -- a power-law spectrum and a flat spectrum, as indicated
by the green lines in Figure~\ref{ffts}. The first often relates
to red noise, while the other spectrum is usually considered
white noise in solar physics and astrophysics
\citep[e.g.][]{Vaughan05,Liang20,Anfinogentov21}. The FFT power
spectra show a clear peak above the 99\% significance level (cyan
lines) at a value that is close to about 20\,seconds, as indicated
by the vertical dotted lines. We would like to note that the FFT
peaks are a bit broad and do not appear above 30\,seconds as
shown by the vertical dashed line. On the other hand, the FFT
spectrum at GOES~0.5\,--\,4\,{\AA} demonstrates double peaks close
to 40\,seconds, which are absent in the GOES~1\,--\,8\,{\AA} flux.
Therefore, only the period of about 20\,seconds for the flare flux
is considered in the current study.

\begin{figure}
\centerline{\includegraphics[width=\textwidth,clip=]{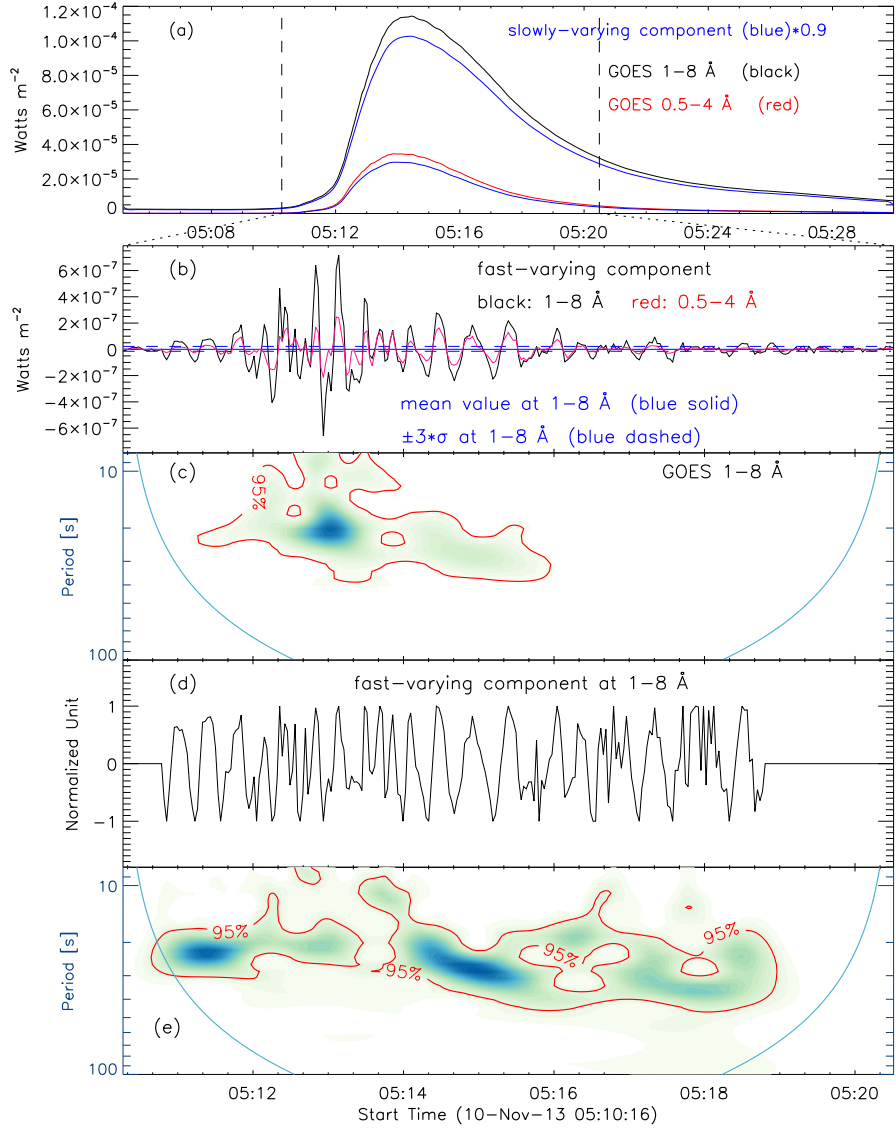}}
\caption{Panel~{\bf a}: SXR light curves in GOES 1\,--\,8\,\AA\
({\it black}), and 0.5\,--\,4\,\AA\ ({\it red}) for the 10 November
2013 solar flare, and their slowly varying components ({\it blue},
after multiplication by 0.9). Panels~{\bf b} and {\bf c}: Rapidly
varying components of GOES SXR fluxes in the time interval between
the {\it two vertical dashed lines} in Panel~{\bf a} and its wavelet power
spectrum. The {\it blue solid and dashed lines} represent the mean and
$\pm$3$\sigma$ at GOES 1\,--\,8\,\AA. Panels~{\bf d} and {\bf e}:
the normalized rapidly varying component at GOES 1\,--\,8\,\AA\ and
its wavelet power spectrum. The {\it red contours} represent a
significance level of 95\,\%. Anything `outside' the {\it light blue
curve} in Panels~{\bf c} and {\bf e} is dubious.} \label{wav1}
\end{figure}

Figure~\ref{wav1}~presents the analysis of periodicity for the
circular-ribbon flare recorded by GOES. Similar to a typical
solar flare, this event exhibits an impulsive increasing and a slow
decaying phase, and no obvious QPP signature is seen in the raw SXR
light curves. This is because the flare QPPs in the GOES SXR flux
often show a very small amplitude, superimposed on a strong
background emission
\citep[e.g.][]{Simoes15,Feng17,Feng20,Kolotkov18,Hayes20}.
Therefore, the strong background emission should be removed from the
original data. In this study, the raw light curve is first
decomposed into a slowly varying component and a rapidly varying
component using the FFT \citep{Ning14,Ning17,Li17,Milligan17}.
Figure~\ref{ffts} suggests that the flare QPP in GOES
1\,--\,8\,\AA\ and 0.5\,--\,4\,\AA\ shows a clear period near
20\,seconds, so a cutoff threshold of 30\,seconds is applied as
marked by the vertical dashed line in Figure~\ref{ffts}. Thus, the
long-term trend is suppressed while the short period can be
highlighted in the wavelet power spectrum
\citep{Kupriyanova10,Kolotkov15,Kashapova20}. Here, the slowly
varying components are multiplied by 0.9 to avoid overlap with
the raw light curves, as shown by the blue curves in
Figure~\ref{wav1}a. The rapidly varying components during the flare
are presented in Panel~b, as outlined by two vertical dashed lines
in Figure~\ref{wav1}a. They are characterized by a series of pulses.
Then the wavelet-analysis method \citep{Torrence98} is applied to
the rapidly varying component seen in GOES 1\,--\,8\,\AA\ as shown
in Panel~c. The wavelet power spectrum confirms a broad range of
enhanced power that is centered at about 20\,seconds, suggesting a
dominant period of about 20\,seconds. However, the enhanced power
focuses on the impulsive phase in the GOES SXR channel, largely due
to the fact that the amplitudes of flare QPPs change with time.
Figure~\ref{wav1}b shows some pulses with a small amplitude during
the flare-decay phase. In order to rule out the noise, we set a
threshold ($\pm$3$\sigma$) to identify the flare QPP pulses. Here,
$\sigma$ represents the standard deviation of the rapidly varying
component. Only those pulse amplitudes that are greater than the
threshold are regarded as valid pulses, as shown by the blue-dashed
lines in Panel~b. Then, each valid pulse of the rapidly varying
component of the GOES 1\,--\,8\,\AA\ is normalized by its pulse
amplitude, as shown in Panel~d. Finally, Panel~e displays the
wavelet power spectrum for the normalized rapidly varying component,
which exhibits a dominant period of roughly 20\,seconds during the
flare time, i.e. from the impulsive phase to the decay phase. This
is consistent with the 20-second period in the FFT power spectra in
Figure~\ref{ffts}.

\begin{figure}
\centerline{\includegraphics[width=\textwidth,clip=]{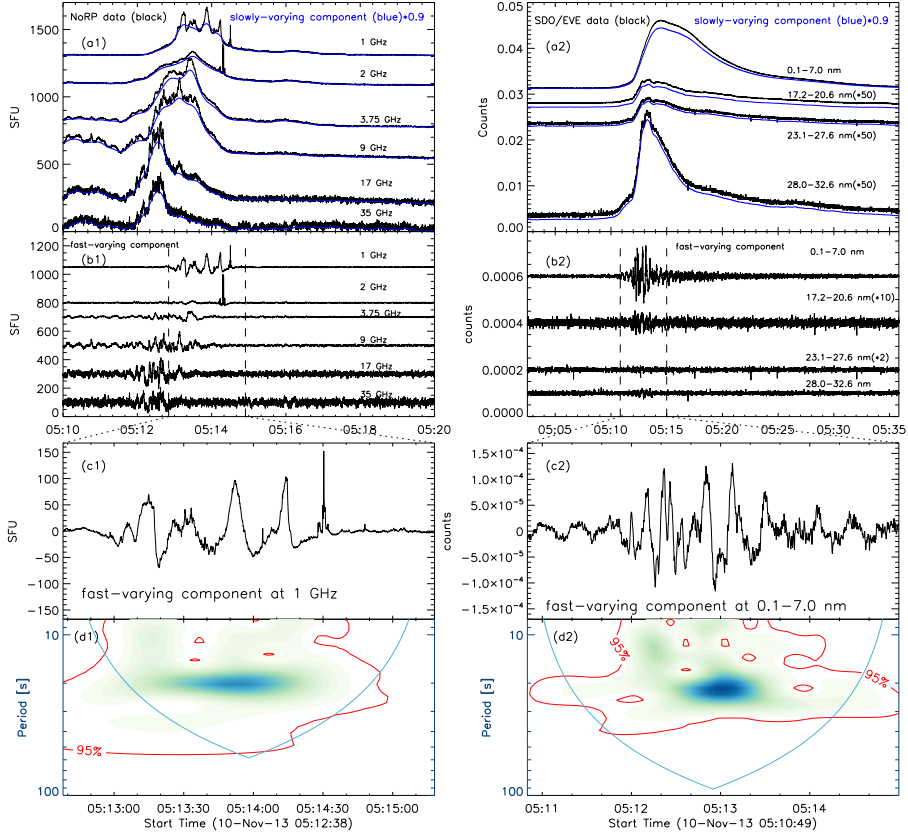}}
\caption{Similar to Figure~\ref{wav1}, the light curves ({\it
black}) detected by NoRP and SDO/EVE, and their slowly
varying components ({\it blue}, after multiplication by 0.9), as well
as the rapidly varying components in the time intervals between the {\it two
vertical dashed lines} and their wavelet power spectra at
NoRP~1\,GHz and SDO/EVE~0.1\,--\,7\,nm. The {\it red
contours} represent a significance level of 95\,\%.} \label{wav2}
\end{figure}

\begin{figure}
\centerline{\includegraphics[width=\textwidth,clip=]{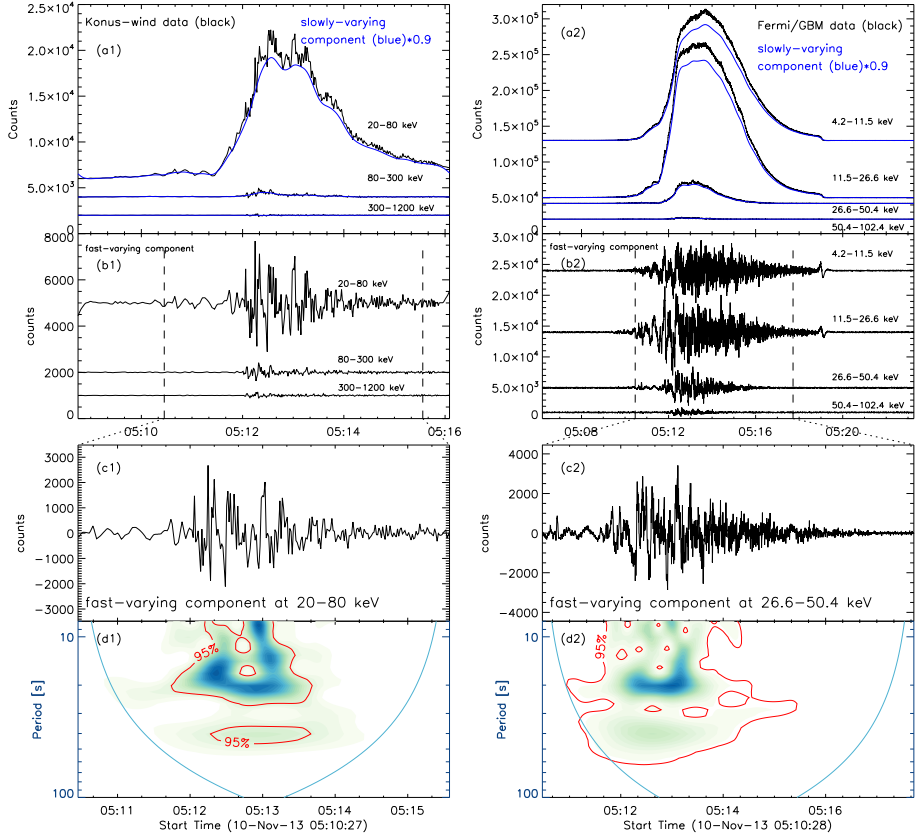}}
\caption{Similar to Figure~\ref{wav1}, the light curves ({\it
black}) detected by {\it Konus-Wind} and {\it Fermi}/GBM,
and their slowly varying components ({\it blue}, after multiplication
by 0.9), as well as the rapidly varying components in the time intervals
between the {\it two vertical dashed lines} and their wavelet power
spectra at KW 20\,--\,80\,keV, and {\it Fermi} 26.6\,--\,50.4\,keV,
respectively. The {\it red contours} represent a significance level
of 95\,\%.} \label{wav3}
\end{figure}

\begin{figure}
\centerline{\includegraphics[width=\textwidth,clip=]{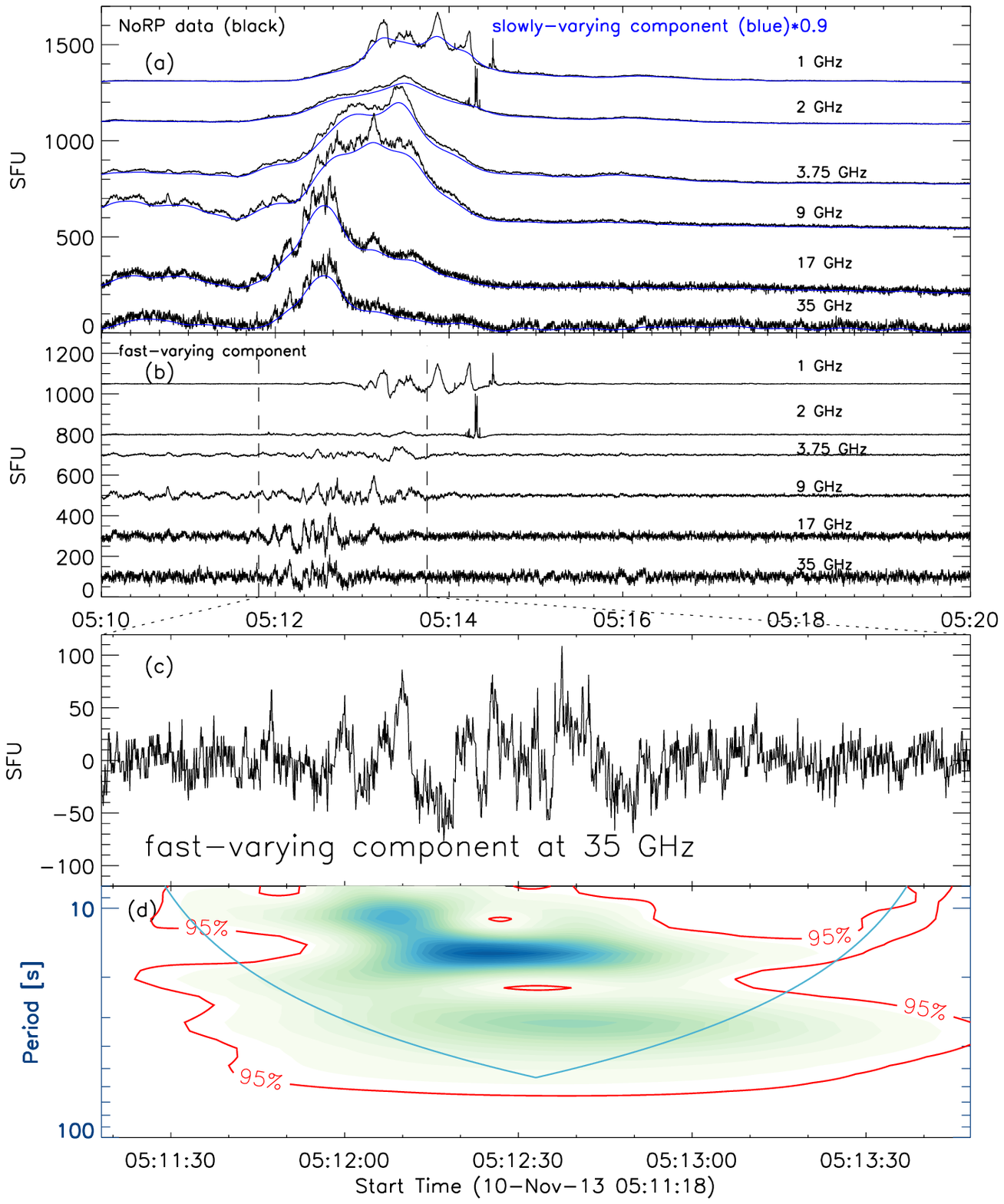}}
\caption{Similar to Figure~\ref{wav1}, the light curves ({\it
black}) detected by the NoRP, and their slowly varying components
({\it blue}, after multiplication by 0.9), as well as the rapidly
varying components in the time intervals between the {\it two vertical dashed
lines} and their wavelet power spectra at NoRP~35\,GHz. The {\it red
contours} represent a significance level of 95\,\%.} \label{wav4}
\end{figure}

In addition to GOES, the X1.1 flare was also observed by NoRP,
RHESSI, {\it Konus-Wind}, SDO/EVE, and {\it Fermi}/GBM in
radio/microwave, X-ray, and EUV ranges, as shown in Table~\ref{tab1}.
Similar to the GOES full-disk flux, any of them detect solar
radiation from the entire Sun. We should point out that RHESSI
provides imaging spectroscopy, but we have focused on the whole disk
(Sun-as-a-star) observations in this study. The RHESSI light curve
is not used here for periodicity tests, because there have been several
changes in the attenuators (see
\textsf{hessi.ssl.berkeley.edu/hessidata/metadata/2013/11/10/hsi\_20131110\_040600\_rate.png}).
Then the same wavelet-analysis method \citep{Torrence98} is used
for the periodicity tests in the microwave, SXR/HXR, and EUV
wavelengths; the detailed results can be seen in
Figures~\ref{wav2}\,--\,\ref{wav4}. The raw light curves (black) are
decomposed into slowly (blue) and rapidly varying components, as
shown in Panels~a1, a2, b1, and b2 in Figures~\ref{wav2} and
\ref{wav3}. The slowly varying components have been multiplied by 0.9 to
avoid overlap with the raw light curves, and some light curves
have been shifted in height, so that they can be displayed clearly
in the same window. Panels~c1 and c2 present the rapidly varying
components observed by NoRP~1\,GHz, ESP~0.1\,--\,7\,nm,
KW~20\,--\,80\,keV, and {\it Fermi}~26.6\,--\,50.4\,keV during the flare,
as indicated by two vertical dashed lines in Panels~b1 and b2. Here,
the NoRP flux at a frequency of 1\,GHz is used to perform the
wavelet analysis because it exhibits the most obvious oscillation
as shown in Figure~\ref{wav2}a1 and b1. The wavelet analysis of SDO/EVE ESP~0.1\,--\,7\,nm is displayed, as it is very
similar to the GOES SXR flux. The other light curves measured by
NoRP and SDO/EVE also show the QPP feature, but they are very weak,
in particular for the EUV flux at ESP~28.0\,--\,32.6\,nm, as shown
in Figure~\ref{wav2}b1. Therefore, the wavelet power spectra in the
NoRP~1\,GHz, ESP~0.1\,--\,7\,nm, KW~20\,--\,80\,keV, and
{\it Fermi}~26.6\,--\,50.4\,keV are presented here, which show that the
pulsation is strongest at a periodicity of 20\,seconds, but the
oscillation period is not exactly 20\,seconds. It is spread over a
broad range of periods, but centered at about 20\,seconds. This is
similar to the previous finding obtained from the GOES SXR flux.
Physically, the microwave emission indicates nonthermal processes at
frequencies above the frequency related to the flux maximum, the
so-called ``peak frequency". In the current case, the microwave
emission at NoRP~35\,GHz is exactly above the ``peak frequency".
Thus, we also present the wavelet analysis method for the microwave
flux at the frequency of NoRP~35\,GHz, as shown in
Figure~\ref{wav4}. Similar to previous results, the wavelet power
spectrum shows an enhanced power over a broad range of periods, and
the strongest power is centered at about 20\,seconds. On the other
hand, the onset time of the 20-second QPP in NoRP~35\,GHz is earlier
than in NoRP~1\,GHz, i.e. one minute earlier.

\subsection{Periodic Jets Triggered by the Circular Ribbon Flare}
\begin{figure}
\centerline{\includegraphics[width=\textwidth,clip=]{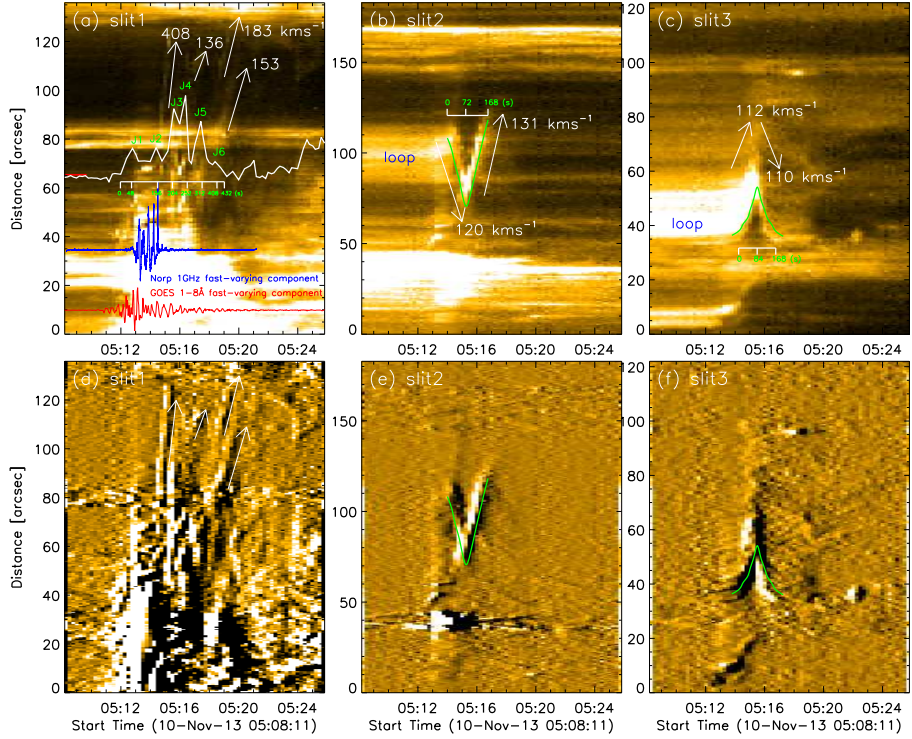}}
\caption{Time--distance images along three slits derived from the
AIA~171\,\AA\ image series ({\bf a}\,--\,{\bf c}) and their
running-difference ({\bf d}\,--\,{\bf f}). The {\it white line} profile in
Panel~{\bf a} is the intensity at the site
$\approx$66$^{\prime\prime}$, as marked by the {\it red tick} on the
left hand. The {\it overplotted blue and red} lines represent the
rapidly varying components in NoRP~1\,GHz and GOES~1\,--\,8\,\AA,
respectively. The {\it white arrows} outline the velocity of the jets. The
{\it green curves} mark the loop perturbation.} \label{tds}
\end{figure}

Figure~\ref{tds} shows the time--distance images along three slits
derived in the AIA~171\,\AA\ image series (a\,--\,c) and their
running-difference images (d\,--\,f). As shown in
Figure~\ref{image}, a faint outer loop appears in the AIA~171\,\AA\
during the X1.1 flare, connecting the closed circular ribbon
and the remote brightening. Here, the slits are selected to cross
the footpoint (S1), the loop apex (S2), and the loop leg (S3) for
the outer loop, as outlined by the white curves in
Figure~\ref{image}b. Panels~a and d plot the time--distance images
along the Slit~1 (S1) at the footpoint, the bright emission at the
bottom is emitted from the closed circular ribbon of the X1.1 flare,
which is accompanied by a series of coronal jets. The coronal jets
are very weak in the original image, but they are much clearer in
the running-difference image, which are characterized by the dark
oblique stripes, as indicated by the white arrows in Panel~d. The
coronal jets exhibit a fast speed that could be as high as
408\,km~s$^{-1}$. The average jet velocity is estimated by the slope
of dark oblique stripes in the running-difference image. The jets occurred
periodically. We plotted the intensity variations (white)
along the 66$^{\prime\prime}$ marked by the red line on Panel a. The
intensity curves show six peaks corresponding to six jets as
J1\,--\,J6 marked on timescale during the time interval of
$\approx$432\,seconds. Thus, the average cadence of the coronal jets is
about 72\,seconds. However, we could not find a similar
quasi-period in the X-ray or microwave flux. The overplotted blue
and red lines are the rapidly varying components in
NoRP~1\,GHz and GOES 1\,--\,8\,\AA\ in Figure~\ref{tds}a.
They both show the short-period oscillation, i.e. a dominant period
of 20\,seconds, which is much shorter than the average cadence of the
coronal jets. Moreover, the 20-second oscillation in GOES and NoRP
occurs in the same interval between about 05:12 UT and 05:15 UT as
the first two jets J1 and J2, while the other jets from J3 to J6
appear after 05:16 UT, when the 20-second oscillation
disappears.

\subsection{Transverse Perturbation of the Outer Loop}
\begin{figure}
\centerline{\includegraphics[width=\textwidth,clip=]{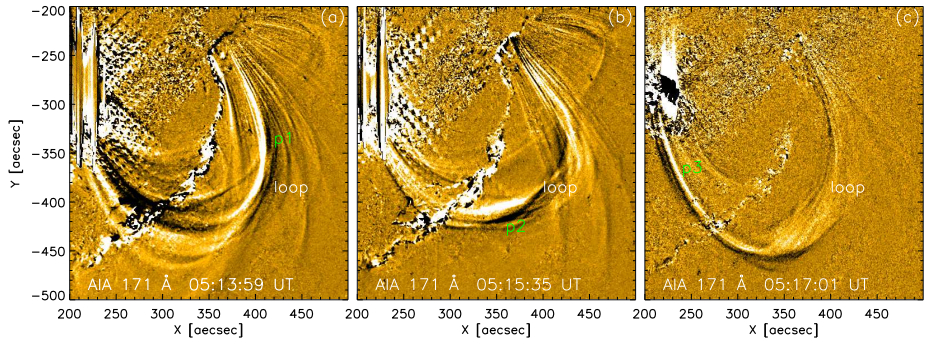}}
\caption{Running-difference images to show the outer loop in
AIA~171\,\AA\ at 05:13:59~UT ({\bf a}), 05:15:35~UT ({\bf b}),
05:17:01~UT ({\bf c}). {\it P1}\,--\,{\it P3} mark the loop apex and two loop
legs.} \label{loops}
\end{figure}

Figure~\ref{tds}b and c present the time--distance images along
slits S2 and S3: the zero of the $y$-axis is marked by the red ``*''
in Figure~\ref{image}b. The outer loop shows the transverse
perturbation consisting of half a cycle only, for instance, it is
displaced from the equilibrium position, reaches maximum
displacement, and returns back to equilibrium without
overshooting. The transverse perturbation has a duration of about
168\,seconds. The onset of the perturbation coincided with the flare
eruption, see also the movie in the AIA~171\,\AA\ channel. The
perturbation amplitude is very large, maybe as high as 20\,Mm,
suggesting a large-amplitude perturbation. It is interesting that
the loop apex (S2) shows an anti-phase perturbation with the loop
leg (S3). In other words, the loop apex shrinks, while the loop
leg expands. Figure~\ref{tds}e and f give their time--distance
images after running-difference. Figure~\ref{loops} shows the
running-difference images in the AIA~171\,\AA\ band. The temporal
cadence to perform the running-difference image is 24\,seconds
and not 12\,seconds. This is because the AIA images have
two different exposure times during this solar flare. Consistent
with the anti-phase perturbations in Figure~\ref{tds}e and f, the
different parts of the outer loop exhibit the various perturbation
directions. For example, at 05:13:59~UT in Figure~\ref{loops}a, the
loop leg of P1 shrinks, while the loop apex of P2 and another
loop leg of P3 expand. At 05:15:35~UT in Figure~\ref{loops}b,
the loop leg of P1 expands, while the loop apex of P2 and
another loop leg of P3 shrink. Such anti-phase perturbations
are possibly the signature of an MHD wave in the outer loop
\citep[see][]{Duckenfield19}.

\section{Conclusion and Discussion}
Based on observations obtained by various instruments in the X-ray,
microwave, and UV range, we investigate flare-related oscillations
during the SOL2013-11-10T05:14 event. SDO/AIA--1600\,\AA\ and
171\,\AA\ bands show that the X1.1 flare is a typical
circular-ribbon flare. The spine-fan magnetic structure derived
from the NLFFF extrapolation model \citep{Wheatland00,Wiegelmann04}
confirmed the magnetic topology specific to the circular-ribbon
flare. For instance, the dome-like structure is consistent with the
closed circular ribbon: A set of overlying magnetic-field lines is
consistent with the faint outer loop seen in the AIA~171\,\AA\ band
image (see Figure~\ref{extr}). Our NLFFF extrapolation agrees with
previous results using the potential magnetic-field extrapolation
\citep[e.g.][]{Meszarosova13}. We found three types of oscillations:
 i) flare QPP with a quasi-period of roughly 20\,seconds in the
SXR/HXR, radio, and EUV channels, ii) periodic jets following the
circular-ribbon flare with a cadence of about 72\,seconds, iii) a
transverse perturbation with a duration of about 168\,seconds in the
faint outer loop.

The FFT and wavelet-analysis methods are applied to determine the
periodicity of the flare fluxes at multiple wavelengths, and
both show an oscillatory period of roughly 20\,seconds. The wavelet
spectra show an enhanced power over a broad range, but the wavelet
power is always centered at the periodicity of around 20\,seconds in
SXR/HXR, radio, and EUV passbands. The GOES SXR flux shows a sharp,
increasing trend during the flare, so the small-amplitude QPPs only
appear clearly during a short time interval, as shown in
Figure~\ref{wav1}c. They could be clearly seen during the whole
flare time after normalization, as shown in Figure~\ref{wav1}e. The
flare QPP with a similar period can also be observed in the X-ray,
radio/microwave, and EUV wavelength ranges recorded by other instruments.
This is similar to previous observations
\citep{Nakariakov10,Milligan17,Kupriyanova20}. The flare QPP with a
broad period range of 10\,--\,83\,seconds has been found in a
circular-ribbon flare \citep{Meszarosova13}, containing our
20-second periodicity. They further identified the wavelet tadpoles
from the radio source and explained them as fast magnetoacoustic
waves propagating in the coronal loop. However, we could not get the
wavelet tadpoles from the wavelet power spectra. The flare QPP with
a dominant period of $\approx$20\,seconds is most likely
triggered by the nonthermal process. Our multi-wavelength
observations suggest the co-existence of thermal and nonthermal
processes during the circular-ribbon flare. The flare QPP is
detected at SXR and EUV wavelengths, which is a thermal process.
It is also seen in the microwave emissions, which obviously
decay after reaching the maximum, indicating nonthermal
gyrosynchrotron emission during the flare
\citep{Dulk85,Warmuth16,Li20c}. Moreover, the flare QPP can also be
observed in the HXR and high-frequency (i.e. NoRP~35\,GHz) microwave
emissions, implying an energy-release process by nonthermal
electrons. The Neupert effect \citep{Neupert68,Ning08,Ning09,Ning10}
is plasma heating as a result of energy released by nonthermal
electrons. So, the oscillations of the accelerated electron emission are
one reason for the oscillations detected in the thermal emissions in this
study.

Similar to previous findings
\citep[e.g.][]{Reid12,Shen12,Zhang16,Lu19}, we have found a series of
coronal jets triggered by the circular-ribbon flare at the footpoint
of the outer loop, and they appear to have a regular and repeated occurrence
with an average periodicity of roughly 72\,seconds. These periodic
jets could be regarded as repeated outflows produced by magnetic
reconnection during circular-ribbon flare
\citep{Chitta17,Shen19,Zhang20}. Therefore, the periodic jets at the
footpoint might be produced by a nonthermal process during the
circular-ribbon flare, i.e. by repetitive magnetic reconnection.
This is consistent with previous findings \citep{Kumar15}, that found
that repeated magnetic reconnection at the footpoint could drive the
periodic acceleration of nonthermal electrons. However, we could not
detect a similar periodicity such as 72\,seconds in flare light
curves in HXR and microwave emissions, both of which are often
related to the nonthermal electron. The dominant period of the flare
QPP is about 20\,seconds, which is far from the periodicity of
coronal jets. So it is difficult to conclude whether the flare QPP has
affected the periodicity jets or whether it is the result of it.

The outer-loop appears to show a transverse perturbation with a
duration of about 168\,seconds during this event. The perturbation is
similar to previous observations of transverse oscillations in
the outer loop of the circular-ribbon flare
\citep[e.g.][]{Zhang20b,Dai21}. However, the perturbation feature
studied here only persists for about half a cycle and then
disappears, so it cannot be regarded as an oscillation or
periodicity. Here, the 168\,seconds are regarded as the perturbation
duration of the loop apex and leg. It is interesting that loop apex
perturbed in anti-phase to one of the loop legs, which would
suggest that the loop perturbation is probably driven and modulated
by a sausage wave \citep{Tian16,Lib20} or a slow magnetoacoustic wave
\citep{Meszarosova13,Kumar15}. It should be pointed out that the
outer-loop perturbation only remains half a cycle, which might be
due to the large perturbation amplitude, since the decay time
depends on the perturbation/oscillation amplitude
\citep[e.g.][]{Goddard16}.

\begin{acks}
The authors thank the reviewer for their valuable comments. We thank
T.~Wiegelmann, Y. Guo, and K. Yang for sharing the NLFFF and vector
magnetic-field analysis codes. We would also acknowledge Prof. Brian Dennis for helping the RHESSI image data.

We thank the teams of SDO/AIA, SDO/HMI, GOES, SDO/EVE, RHESSI, {\it
Konus-Wind}, {\it Fermi}, NoRP, and NoRH for their open-data-use
policy. Data courtesy of NASA/SDO and the AIA science team. The {\it
Ramaty High Energy Solar Spectroscopic Imager} is a NASA Small
Explorer mission. The {\it Nobeyama Radioheliograph} and {\it
Nobeyama Radio Polarimeters} are operated by the Nobeyama Solar
Radio Observatory, NAOJ/NINS.

This work is supported by NSFC under grants 12073081, 12003072,
11973092, 11790302, 11729301, U1731241, as well as CAS Strategic
Pioneer Program on Space Science, Grant No. XDA15052200,
XDA15320301. Y.~Wang also acknowledges the Youth Fund of JiangSu No.
BK20191108. D.~Li is also supported by the Surface Project of
Jiangsu No. BK20211402. The Laboratory No. is 2010DP173032.
\end{acks}

\section*{Declarations}
\textbf{Conflict of Interest}
The authors declare that they have no conflict of interest.

\bibliographystyle{spr-mp-sola}

\bibliography{SOLA-D-21-089_R3_LI_references}
\IfFileExists{\jobname.bbl}{} {\typeout{}
\typeout{****************************************************}
\typeout{****************************************************}
\typeout{** Please run "bibtex \jobname" to obtain} \typeout{**
the bibliography and then re-run LaTeX} \typeout{** twice to fix
the references !}
\typeout{****************************************************}
\typeout{****************************************************}
\typeout{}}

\end{article}
\end{document}